\documentclass[12pt]{article}
 \def\lddots{\mathinner{\mkern1mu\raise1pt\hbox{.}\mkern2mu
    \raise4pt\hbox{.}\mkern2mu\raise7pt\vbox{\kern7pt\hbox{.}}\mkern1mu}}
\makeatletter
\def\numberbysection{\@addtoreset{equation}{section}
 	\def\theequation{\thesection.\arabic{equation}}}
\makeatother

\numberbysection

\newcommand{\be}{\begin{eqnarray}}
\newcommand{\ee}{\end{eqnarray}}
\newcommand{\non}{\nonumber}
\newcommand{\n}{\ensuremath{\mathcal{N}}}
\newcommand{\tr}{\mathop{\rm tr}\nolimits}

\begin{document}

\begin{titlepage}
\strut\hfill DTP/00/46
\vspace{.5in}
\begin{center}

\LARGE Quantum spin chain with ``soliton non-preserving'' boundary conditions\\[1.0in]
\large Anastasia Doikou\\[0.8in]
\large Department of Mathematical Sciences, University of Durham,\\[0.2in]
\large South Road, DH1 3LE, Durham, United Kingdom\\

\end{center}

\vspace{.5in}

\begin{abstract}

 We consider the case of an integrable quantum spin chain with ``soliton non-preserving'' boundary conditions. This is the first time that such boundary conditions have been considered in the spin chain framework. We construct the transfer matrix of the model, we study its symmetry and we find explicit expressions for its eigenvalues. Moreover, we derive a new set of Bethe ansatz equations by means of the analytical Bethe ansatz method. 

\end{abstract}

\end{titlepage}

\section{Introduction}

 So far, quantum spin chains with ``soliton preserving'' boundary conditions have been studied \cite{DVGR3}-\cite{dn/duality}. However, there exists another type of boundary conditions, namely the ``soliton non-preserving''. These conditions are basically known in affine Toda field theories \cite{delius1}-\cite{gand}, although there is already a hint of such boundary conditions in the prototype paper of Sklyanin \cite{sklyanin}, which is further clarified by Delius in \cite{delius1}. It is important to mention that in affine Toda field theories only the ``soliton non-preserving'' boundary conditions have been studied \cite{gand}, \cite{corr}. It is still an open question what the ``soliton preserving'' boundary conditions are in these theories.

In this work we construct the open spin chain with the ``new'' boundary conditions, we show that the model is integrable, we study its symmetry, and evidently, we solve it by means of the analytical Bethe ansatz method \cite{reshe}-\cite{amn/spectrum}. This is the first time that such boundary conditions have been considered in the spin chain framework.

To describe the model it is necessary to introduce the basic constructing elements, namely, the $R$ and $K$ matrices.

 The $R$ matrix, which is a solution of the Yang-Baxter equation
\be
R_{12}(\lambda_{1} - \lambda_{2})\ R_{1 3}(\lambda_{1})\ R_{23}(\lambda_{2}) = R_{23}(\lambda_{2})\ R_{1 3}(\lambda_{1})\ R_{1 2}(\lambda_{1} - \lambda_{2})\
\label{YBE}
\ee
(see, e.g., \cite{QISM}). 

Here, we focus on the special case of the $SU(3)$ invariant $R$ matrix
\cite{yang}
\be
R_{12}(\lambda)_{j j \,, j j} &=& {(\lambda + i)} \,, \non \\
R_{12}(\lambda)_{j k \,, j k} &=& { \lambda} \,, \qquad j \ne k
\,, \non \\
R_{12}(\lambda)_{j k \,, k j} &=& {i}\,, \qquad j \ne k \,, \non \\
& & 1 \le j \,, k \le 3 \,.
\label{Rmatrix}
\ee
We also need to introduce the $R$ matrix that involves different representations of $SU(3)$ \cite{VEWO}, \cite{abad}, in particular, $3$ and $\bar 3$ (see also \cite{doikou}). This matrix is given by crossing \cite{ogiev}-\cite{delius}
\be
R_{\bar 12}(\lambda) = V_{1}\ R_{12}(-\lambda - \rho)^{t_{2}}\ V_{1} = V_{2}^{t_{2}}\ R_{12}(-\lambda - \rho)^{t_{1}}\ V_{2}^{t_{2}} \,,
\label{prop4}
\ee
where $V^{2}=1$. 
Note that, 
\be
R_{\bar i j}(\lambda)= R_{i \bar j}(\lambda) \equiv \bar R_{ij}(\lambda)\,, \qquad R_{ \bar i \bar j}(\lambda) = R_{ij}(\lambda)\,.
\label{note}
\ee
The $\bar R$ matrix is also a solution of  the Yang-Baxter equation
\be
\bar R_{ 1 2}(\lambda_{1} - \lambda_{2})\ \bar R_{1 3}(\lambda_{1})\ R_{23}(\lambda_{2}) = R_{23}(\lambda_{2})\ \bar R_{1 3}(\lambda_{1})\ \bar R_{1 2}(\lambda_{1} - \lambda_{2})\,.
\label{YBE2}
\ee
The matrices $K^{-}$, and $K^{+}$ which are solutions
of the boundary Yang-Baxter equation \cite{cherednik}, \cite{gand}
\be
R_{12}(\lambda_{1}-\lambda_{2})\ K_{1}^{-}(\lambda_{1})\
\bar R_{2 1}(\lambda_{1}+\lambda_{2})\ K_{2}^{-}(\lambda_{2}) \non \\
= K_{2}^{-}(\lambda_{2})\ \bar R_{12}(\lambda_{1}+\lambda_{2})\
K_{1}^{-}(\lambda_{1})\ R_{21}(\lambda_{1}-\lambda_{2}) \,,
\label{boundaryYB1}
\ee
and,
\be
R_{12}(-\lambda_{1}+\lambda_{2})\ K_{1}^{+}(\lambda_{1})^{t_{1}}\
\bar R_{2 1}(-\lambda_{1}-\lambda_{2}-2\rho)\ K_{2}^{+}(\lambda_{2})^{t_{2}} \non \\
= K_{2}^{+}(\lambda_{2})^{t_{2}}\ \bar R_{12}(-\lambda_{1}-\lambda_{2}-2\rho)\
 K_{1}^{+}(\lambda_{1})^{t_{1}}\ R_{21}(-\lambda_{1}+\lambda_{2}) \,,
\label{boundaryYB2}
\ee
where $\rho = {3i \over 2}$. We can consider that the $K_{i}$ matrix describes the reflection of a soliton with the boundary which comes back as an anti-soliton (see also \cite{gand}). 

 It is a natural choice to consider the following alternating spin chain \cite{VEWO}, \cite{abad}, which leads to a local Hamiltonian. The corresponding transfer matrix $t(\lambda)$ for the open chain of $2N$ sites with ``soliton non-preserving'' boundary conditions is (see also e.g., \cite{sklyanin}, \cite{mn/nonsymmetric})
\be
t(\lambda) = \tr_{0}  K_{0}^{+}(\lambda)\
T_{0}(\lambda)\  K^{-}_{0}(\lambda)\ \hat T_{\bar 0}(\lambda)\,,
\label{transfer1}
\ee
where $\tr_{0}$ denotes trace over the ``auxiliary space'' 0,
$T_{0}(\lambda)$ is the monodromy matrix,
\be
T_{0}(\lambda) = R_{0 2N}(\lambda) \bar R_{0 2N-1}(\lambda)\cdots R_{0 2}(\lambda) \bar R_{0 1 }(\lambda) \,, \non\\ \hat T_{\bar 0}(\lambda) = R_{1 0}(\lambda) \bar R_{2 0}(\lambda)\cdots R_{2N-1 0}(\lambda) \bar R_{2N 0}(\lambda) \,,
\label{hatmonodromy}
\ee
 We can change the auxiliary space to its conjugate and then we obtain the $\bar t(\lambda)$ matrix which satisfies, for $K^{\pm}(\lambda) = 1$
\be
\bar t(\lambda) = t(\lambda)^{t}\,.
\ee
In particular,
\be
\bar t(\lambda) = \tr_{0}  K_{\bar 0}^{+}(\lambda)\
T_{\bar 0}(\lambda)\  K^{-}_{\bar 0}(\lambda)\ \hat T_{0}(\lambda)\,,
\label{transfer2}
\ee
with
\be
T_{\bar 0}(\lambda) = \bar R_{0 2N}(\lambda) R_{0 2N-1}(\lambda)\cdots \bar R_{0 2}(\lambda) R_{0 1 }(\lambda) \,, \non\\ \hat T_{0}(\lambda) = \bar R_{1 0}(\lambda)  R_{2 0}(\lambda)\cdots \bar R_{2N-1 0}(\lambda)  R_{2N 0}(\lambda) \,,
\label{hatmonodromyp}
\ee
(we usually suppress the ``quantum-space'' subscripts
$1 \,, \ldots \,, N$).  One can observe the alternation between $R$ and $\bar R$ in (\ref{hatmonodromy}) and (\ref{hatmonodromyp}). In particular for the monodromy matrix $T_{0}$ we see that in even sites there exists the $R$ matrix whereas in the odd sites the $\bar R$ matrix acts. The situation is exactly the opposite for the $\hat T_{\bar 0}$ matrix. In fact, 
\be
\hat T_{a}(\lambda) = T_{a}^{-1}(-\lambda)\,,
\ee
where $a$ can be $0$ or $\bar 0$. In the above definitions of the monodromy matrices we used the equations (\ref{note}).

The transfer matrix satisfies the commutativity property
\be
\left[ t(\lambda)\,, t(\lambda') \right] = 0 \,.
\label{commutativity}
\ee
$\bar t$ also obeys the commutativity property,
\be
\left[ \bar t(\lambda)\,, \bar t(\lambda') \right] = 0 \,,
\label{commutativityp}
\ee
moreover
\be
\left [\bar t(\lambda), t(\lambda')\right] = 0\,.
\label{commutativity3}
\ee
We give a detailed prof of (\ref{commutativity}), (\ref{commutativityp}), (\ref{commutativity3}) in A Appendix.
The corresponding open spin chain Hamiltonian $\cal H$ is
\be
{\cal H} \propto {d\over d \lambda}t(\lambda) \bar t(\lambda) \Big\vert_{\lambda=0} \,.
\label{hamilt}
\ee
It is necessary to consider the product of both transfer matrices in order to obtain a local theory. One can show that this Hamiltonian is indeed local with terms that describe interaction up to four neighbours, see B Appendix.

The outline of this paper is as follows: in the next section we briefly discuss about the crossing symmetry of the transfer matrix and the fusion for the $K$ matrices and the transfer matrix. In section three we study the asymptotic behaviour and  the symmetry of the transfer matrix. We show that, although we build the chain using the $SU(3)$ invariant $R$ matrix, the model has $SO(3)$ symmetry. In the following section we find the exact expressions for the transfer matrix eigenvalues and we also deduce a completely new set of Bethe ansatz equations via the analytical Bethe ansatz method. Finally, in the last section we review the results of this work and we also discuss some of our future goals.

\section{Crossing and fusion}

In this section we basically review known ideas about the crossing and the fusion procedure for the $R$ and $K$ matrices ( see e.g., \cite{karowski}, \cite{mn/fusion}, \cite{doikou}).
We can prove (see also \cite{mn/anal}) that the transfer matrix satisfies the crossing symmetry. To do this we need the next identity
\be
{\cal P}_{12}^{t_{2}}\ \bar R_{2 1}(\lambda)^{t_{1}} = \bar R_{21}(\lambda)^{t_{1}}\
 {\cal P}_{12}^{t_{2}} \,,
\label{bound}
\ee
where $\cal P$ is the permutation operator.
The last equation follows from the reflection equation (\ref{boundaryYB1}) for $\lambda_{1} - \lambda_{2} = -\rho$.
which follows from the reflection equation (\ref{boundaryYB1}) for $\lambda_{1} - \lambda_{2} = -\rho$.
Then we can show for the transfer matrix that
\be
t(\lambda) = t(-\lambda -\rho) \,.
\label{cross}
\ee
Indeed, the transfer matrix does have crossing symmetry.

The fused $R$ matrices are known (see e.g., \cite{doikou}). However we still need to fuse the $K$ matrices.
We consider the following reflection equation for $\lambda_{1}-\lambda_{2}=-\rho$,
\be
\bar R_{12}(\lambda_{1}-\lambda_{2})\ K_{\bar1}^{-}(\lambda_{1})\
R_{21}(\lambda_{1}+\lambda_{2})\ K_{2}^{-}(\lambda_{2}) \non \\
= K_{2}^{-}(\lambda_{2})\ R_{12}(\lambda_{1}+\lambda_{2})\
K_{\bar 1}^{-}(\lambda_{1})\ \bar R_{21}(\lambda_{1}-\lambda_{2}) \,,
\label{boundaryYB1p}
\ee
then the fused $K$ matrices are given by
\be
& &K^{-}_{<\bar1 2>}(\lambda) = P_{\bar1 2}^{+}\ K^{-}_{\bar1}(\lambda)\ R_{21}(2 \lambda + \rho)\ K^{-}_{2}(\lambda+\rho)\ P_{2 \bar1}^{+} \,, \non \\
& &K_{<\bar1 2>}^{+}(\lambda)^{t_{12}} = P_{2 \bar1 }^{+}\ K_{\bar1}^{+}(\lambda)^{t_{1}}\ R_{21}(-2 \lambda -3 \rho)\ K_{2}^{+}(\lambda+\rho)^{t_{2}}\ P_{ \bar1 2 }^{+} \,.
\ee
where
\be
P_{\bar12}^{+} = 1-{1\over3} \bar R_{12}(-\rho)
\ee
is a projector to an eight dimensional subspace (see also \cite{doikou}) (${1\over3} \bar R_{12}(-\rho)$ is a projector to an one dimensional subspace).
Analogously, we obtain the $K_{<1 \bar 2>}(\lambda)$ matrices. The above $K$ matrices obey generalised reflection equations (see e.g., \cite{mn/fusion}, \cite{doikou}).
One can show, for the case that $K^{\pm}(\lambda) = 1$, the fused transfer matrix is (see e.g., \cite{mn/fusion})
\be
\hat t(\lambda) = \zeta'(2\lambda+2\rho)\ \bar t(\lambda)\ t(\lambda + \rho) - \zeta(\lambda+\rho)^{N} \zeta'(\lambda+\rho)^{N} g(2\lambda +\rho)g(-2 \lambda - 3\rho)\,, 
\label{fusiont}
\ee
where we define,
\be
g(\lambda) = \lambda + i\,, \qquad \zeta(\lambda) = (\lambda+i)(-\lambda+i)\,, \qquad \zeta'(\lambda)=(\lambda+\rho)(-\lambda+\rho)\,.
\ee
Note that we obtain one equation from fusion whereas in \cite{doikou} we end up with two such equations.

\section{The symmetry of the transfer matrix}

Here, we study the symmetry of the transfer matrix for the alternating spin chain. To do so it is necessary to derive the asymptotic behaviour of the monodromy matrix. The
asymptotic behaviour of the $R$, $\bar R$ matrices for $\lambda \rightarrow \infty$ follows from (\ref{Rmatrix}), (\ref{prop4})
\be
R_{0k}(\lambda) & \sim & \lambda(I + {i\over \lambda} \left(
           \begin{array}{ccc}
            S_{1,k}      &J^{-}_{1,k}  &J^{-}_{3,k}  \\
            J^{+}_{1,k}  & S_{2,k}     &J^{-}_{2,k}   \\               
            J^{+}_{3,k}  &J^{+}_{2,k}  &S_{3,k}\\ 
                                                                         \end{array}\right))\,,\non\\
\bar R_{0k}(\lambda) & \sim & - \lambda(I + {3i \over 2 \lambda}I -  {i \over \lambda} \left
           (\begin{array}{ccc}
             S_{3,k}      &-J^{-}_{2,k}  &J^{-}_{3,k}                \\
             -J^{+}_{2,k} & S_{2,k}      &-J^{-}_{1,k}              \\                      J^{+}_{3,k}  &-J^{+}_{1,k}  &S_{1,k} \\
                                                                         \end{array} \right)).\
\label{asympt}
\ee 
 The matrix elements are:
\be
S_{i}     &=& e_{i,i}\,, \qquad i=1,2,3\,, \non\\
J^{+}_{i} &=& e_{i,i+1}\,, \qquad
J^{-}_{i} = e_{i+1,i}\,, \qquad
 i=1,2\,, \non\\
J^{+}_{3} &=& e_{1,3}\,,\qquad
J^{-}_{3} = e_{3,1}\,,
\ee 
with,
\be
(e_{i,j})_{kl} = \delta_{ik} \delta_{jl}\
\label{e}
\ee
The leading asymptotic behaviour of the monodromy matrix is given by
\be
T^{+}  & \sim & (-)^{{N \over 2}} \lambda^N(I+{3Ni \over 2 \lambda}I+ {i\over \lambda} \left
  (\begin{array}{ccc}
 {\cal S}_{1}^{e}-{\cal S}_{3}^{o}&{\cal J}_{1}^{- e}+{\cal J}_{2}^{- o} &{\cal J}_{3}^{-e}-{\cal J}_{3}^{-o}\\        
{\cal J}_{1}^{+e}+{\cal J}_{2}^{+o}&{\cal S}_{2}^{e}-{\cal S}_{2}^{o}&{\cal J}_{2}^{- e}+{\cal J}_{1}^{- o}\\    
{\cal J}_{3}^{+e}-{\cal J}_{3}^{+o}&{\cal J}_{2}^{+e}+{\cal J}_{1}^{+o}  &{\cal S}_{3}^{e}-{\cal S}_{1}^{o}\\ 
                                                                         \end{array}\right))\,,\non\\
\hat T^{+}  & \sim & (-)^{{N \over 2}} \lambda^{N}(I + {3Ni \over 2\lambda}I + {i \over \lambda} \left
  (\begin{array}{ccc}
 {\cal S}_{1}^{o}-{\cal S}_{3}^{e}&{\cal J}_{1}^{- o}+{\cal J}_{2}^{-e} &{\cal J}_{3}^{-o}-{\cal J}_{3}^{-e}\\        
{\cal J}_{1}^{+o}+{\cal J}_{2}^{+e}&{\cal S}_{2}^{o}-{\cal S}_{2}^{e}&{\cal J}_{2}^{- o}+{\cal J}_{1}^{- e}\\    
{\cal J}_{3}^{+o}-{\cal J}_{3}^{+e}&{\cal J}_{2}^{+o}+{\cal J}_{1}^{+e}  &{\cal S}_{3}^{o}-{\cal S}_{1}^{e}\\                                                                           \end{array}\right)) \,,
\label{asymptT}
\ee
the superscripts $e$ and $o$ refer to the sum over even and odd sites of the chain
respectively, namely,
\be
{\cal S}_{i}^{r}  = \sum_{k=[r]}S_{i,k}\, \qquad  {\cal J}_{i}^{\pm r} = \sum_{k=[r]}J_{i,k}^{\pm}\,, \qquad i=1,2,3\,, 
 \label{quant}
\ee
$r$ can be even or odd.
To determine the symmetry of the transfer matrix we need the asymptotic behaviour of the following product
\be
T^{+} \hat T^{+}   \sim  \lambda^{2N}(I + {3Ni \over \lambda}I + {i \over \lambda} \left
   (\begin{array}{ccc}
   {\cal S}    &{\cal J}^{-}  &0\\        
   {\cal J}^{+} &0             &{\cal J}^{-}\\    
    0           &{\cal J}^{+}  &-{\cal S}\\                                                                           \end{array}\right)) \,,
\label{asymptTT}
\ee
where,
\be 
{\cal S} &=& {\cal S}_{1}-{\cal S}_{3}\,, \qquad {\cal J}^{\pm}={\cal J}_{1}^{\pm}+{\cal J}_{2}^{\pm}\,, 
\ee
 are the generators of $SO(3)$, and
\be
{\cal S}_{i} &=& {\cal S}_{i}^{e}+{\cal S}_{i}^{o} \,, \qquad {\cal J}_{i}^{\pm} = {\cal J}_{i}^{\pm e} + {\cal J}_{i}^{\pm o}\,.
\ee
We define the following operator which has a structure similar to the transfer matrix,
\be
\tau = \tr_{0} P T^{+} \hat T^{+}
\ee
where $P$ can be $S$, $J^{\pm}$ and projects out the corresponding generators from the (\ref{asymptTT}). One can prove (see also \cite{dn/duality}) the commutation relation
\be
\left[ t(\lambda)\,, \tau \right] = 0 \,.
\label{commut}
\ee
 Similarly, one can show that  $\bar t(\lambda)$ commutes with $\tau$, therefore the Hamiltonian (\ref{hamilt}) commutes with $\tau$ as well. It is manifest from the equation (\ref{commut}) that the transfer matrix ($\bar t(\lambda)$ as well) has $SO(3)$ symmetry.
 Even though the result seems ``bizarre'', it is somehow expected if we consider that $SO(3)$ is a subalgebra of $SU(3)$ invariant under charge conjugation. Remember that we constructed the spin chain which involves the 3 and $\bar 3$ representations of $SU(3)$ in both quantum and auxiliary spaces.
Moreover, it is essential for the following to determine the asymptotic behaviour of the transfer matrix eigenvalue which is given by
\be
t(\lambda) \sim \lambda^{2N}(3 + {9Ni \over  \lambda})I
\label{asymt}
\ee
where $I$ is the $3 \times 3$ unit matrix.

\section{Bethe ansatz equations}

We can use the results of the previous sections in order to deduce the Bethe ansatz equations for the spin chain. First, we have to derive a reference state, namely the pseudo-vacuum.
We consider the state with all spins up i.e.,

\be
|\Lambda^{(0)} \rangle =  \bigotimes_{k=1}^{N} |+ \rangle_{(k)}\,,
\label{state}
\ee
 this is annihilated by ${\cal J}^{+}$ where (we suppress the $(k)$ index)
\be
|+ \rangle = \left (\begin{array}{c}
                     1 \\
                     0  \\ 
                     0  \\
                       \end{array} \right)\,.
\label{col}
\ee
This is an eigenstate of the transfer matrix. The action of the $R$, $\bar R$ matrices on the $|+ \rangle$ ($\langle +|$) state gives upper (lower) triangular matrices.
Consequently, the action of the monodromy matrix on the pseudo-vacuum gives also upper (lower) triangular matrices (see also \cite{doikou}).
We find that the transfer matrix eigenvalue for the pseudo-vacuum state, after some tedious calculations, is
\be
\Lambda^{(0)}(\lambda)= (a(\lambda) \bar b(\lambda)) ^{2N} {2 \lambda + {i \over 2} \over 2\lambda+{3i \over 2}} + (b(\lambda) \bar b(\lambda))^{2N} + (\bar a(\lambda)b(\lambda))^{2N}{2\lambda+ {5i \over 2} \over 2\lambda+{3i \over 2}}\,.
\ee
 Because of the $SO(3)$ symmetry of the transfer matrix there exist simultaneous eigenstates of $M={1\over 2}(2N-S)$ and the transfer matrix, namely,
\be
M|\Lambda^{(m)} \rangle \ = m|\Lambda^{(m)} \rangle \,, \qquad t(\lambda)|\Lambda^{(m)} \rangle\ = \Lambda^{(m)}(\lambda) |\Lambda^{(m)} \rangle\,.
\label{eigen}
\ee
We assume that a general eigenvalue has the form of a ``dressed'' pseudo-vacuum  eigenvalue i.e.,
\be
\Lambda^{(m)}(\lambda) =(a(\lambda) \bar b(\lambda)) ^{2N} {2 \lambda + {i \over 2} \over 2\lambda+{3i \over 2}}A_{1}(\lambda) + (b(\lambda) \bar b(\lambda))^{2N}A_{2}(\lambda) + (\bar a(\lambda)b(\lambda))^{2N}{ 2\lambda+ {5i \over 2} \over 2\lambda+{3i \over 2}}A_{3}(\lambda) \,.
\ee
Our task is to find explicit expressions for the $A_{i}(\lambda)$. We consider all the conditions we derived previously.
The asymptotic behaviour of the transfer matrix (\ref{asymt}) gives the following condition for $\lambda \rightarrow \infty$
\be
\sum_{i=1}^{3}A_{i}(\lambda) \rightarrow 3\,.
\ee
The fusion equation (\ref{fusiont}) gives us conditions involving $A_{1}(\lambda)$, $A_{3}(\lambda)$, namely,
\be
A_{1}(\lambda + \rho) A_{3}(\lambda) = 1\,. 
\ee
The crossing symmetry of the transfer matrix (\ref{cross}) provides further restrictions among the dressing functions i.e.,
\be
A_{3}(-\lambda - \rho) =  A_{1}(\lambda)\,, \qquad  A_{2}(\lambda) = A_{2}(-\lambda - \rho)\,.
\label{crosp}
\ee
The last two equations combined give
\be
A_{1}(\lambda)A_{1}(-\lambda) = 1\,.
\ee
 Moreover, for $\lambda= - i$ the $R$ matrix degenerates to a projector onto a three dimensional subspace. Thus, we can obtain another equation that involves $A_{1}(\lambda)$ and $A_{2}(\lambda)$ (see also \cite{reshe}), namely, 
\be
A_{2}(\lambda)A_{1}(\lambda+i)=A_{1}(\lambda+{i \over2})\,.
\ee
 Finally, we require $A_{2}(\lambda)$ to have the same poles with $A_{1}(\lambda)$ and $A_{3}(\lambda)$. Considering all the above conditions together we find that
\be
A_{1}(\lambda) =  \prod_{j=1}^{m} {\lambda+\lambda_{j}-{i\over2}\over \lambda+ \lambda_{j} +{i\over2}}{\lambda-\lambda_{j}-{i\over2}\over \lambda- \lambda_{j} +{i\over2}}\,, 
\ee
\be
A_{2}(\lambda) =  \prod_{j=1}^{m} {\lambda+\lambda_{j}+{3i\over2}\over \lambda+ \lambda_{j} +{i\over2}}{\lambda-\lambda_{j}+{3i\over2}\over \lambda- \lambda_{j} +{i\over2}} {\lambda+\lambda_{j}\over \lambda+ \lambda_{j} +i}{\lambda-\lambda_{j}\over \lambda- \lambda_{j} +i}\,,
\ee
\be
A_{3}(\lambda) =  \prod_{j=1}^{m} {\lambda+\lambda_{j}+2i\over \lambda+ \lambda_{j} +i}{\lambda-\lambda_{j}+2i \over \lambda - \lambda_{j} +i}\,.
\ee
We can check that the above functions indeed satisfy all the necessary properties. Finally, the analyticity of the eigenvalues (the poles must vanish) provides the Bethe ansatz equations
\be
e_{1}(\lambda_{i})^{2N}e_{-1}(2\lambda_{i}) = -\prod_{j=1}^{m} e_{2}(\lambda_{i} - \lambda_{j})\ e_{2}(\lambda_{i} + \lambda_{j})\  e_{-1}(\lambda_{i} - \lambda_{j})\ e_{-1}(\lambda_{i} + \lambda_{j})\,,
\label{BAE}
\ee
where we have defined $e_{n}(\lambda)$ as
\be
e_{n}(\lambda) = {\lambda + {in \over 2} \over \lambda - {in \over 2}}\,.
\label{DEF}
\ee

Notice that we obtain a completely new set of Bethe equations starting with the known $SU(3)$ invariant $R$ matrix. At  this point we can make the following interesting observation. Consider the Bethe ansatz equations for the alternating spin chain with periodic boundary conditions, with $2N$ sites \cite{abad}

\be
e_{1}(\lambda_{i}^{(1)})^{N_{0}} &=&  \prod_{i \ne j=1}^{m_{1}} e_{2}(\lambda_{i}^{(1)} - \lambda_{j}^{(1)})\ \prod_{j=1}^{m_{2}} e_{-1}(\lambda_{i}^{(1)} - \lambda_{j}^{(2)})\,, \non\\
e_{1}(\lambda_{i}^{(2)})^{N_{0}^{*}} &=& \prod_{i \ne j=1}^{m_{2}} e_{2}(\lambda_{i}^{(2)} - \lambda_{j}^{(2)})\ \prod_{j=1}^{m_{1}}  e_{-1}(\lambda_{i}^{(2)} - \lambda_{j}^{(1)})\,,
\label{BAEp}
\ee
($N_{0}+N_{0}^{*}=2N$). For the special case that $N_{0}= N_{0}^{*}= N$, $m_{1} = m_{2}$ and $\lambda_{j}^{(1)} = \lambda_{j}^{(2)}$, the previous equations become
\be
e_{1}(\lambda_{i})^{N} = -\prod_{j=1}^{m} e_{2}(\lambda_{i} - \lambda_{j})\ e_{-1}(\lambda_{i} - \lambda_{j})\,.
\ee

The last equations are exactly ``halved'' compared to (\ref{BAE}). For the moment we do not have any satisfactory explanation about the significance of this coincidence.

Our results can be probably generalized for the spin chain constructed by the $SU(\n)$ invariant $R$ matrix. We expect a reduced symmetry for the general case as well.

\section{Discussion}

We constructed a quantum spin chain with ``soliton non-preserving'' boundary conditions. Although we started with the $SU(3)$ invariant $R$ matrix, we showed that the model has $SO(3)$ invariance (\ref{commut}). We used this symmetry to find the spectrum of the transfer matrix and we also deduced the Bethe ansatz equations (\ref{BAE}) via the analytical Bethe ansatz method. It would be of great interest to study the trigonometric case. Hopefully, one can find diagonal solutions for the $K$ matrices and solve the trigonometric open spin chain. The interesting aspect for the trigonometric case is that one can possibly relate the lattice model with some boundary field theory. Indeed, we know that e.g., the critical periodic $A_{\n-1}^{(1)}$ spin chain can be regarded as a discretisation of the corresponding affine Toda field theory \cite{zinn}. Finally, one can presumably generalize the above construction using any $SU(\n)$ invariant $R$ matrix. We hope to report on these issue!
s in a future work \cite{new}.

\section{Acknowledgments}
I am grateful to E. Corrigan, G.W. Delius, and R.I. Nepomechie for helpful discussions. This work was supported by the European Commission under the TMR Network ``Integrability, non-perturbative effects, and symmetry in quantum field theory'', contract number FMRX-CT96-0012.

\appendix 

\section{Appendix}
In this section we are going to prove the integrability of the model, namely the commutation relations (\ref{commutativity}), (\ref{commutativityp}) for the transfer matrices. 
We define the following operator which originally introduced by Sklyanin \cite{sklyanin},
\be
{\cal T}_{0}^{-}(\lambda) = T_{0}(\lambda)K_{0}^{-}(\lambda) \hat T_{\bar 0}(\lambda)\,.
\label{oper}
\ee
As we already mentioned in the introduction the $K^{-}(\lambda)$ matrix satisfies the reflection equation (\ref{boundaryYB1}) therefore the ${\cal T}_{0}^{-}(\lambda)$ operator obeys the fundamental relation
\be
R_{12}(\lambda_{1}-\lambda_{2})\ {\cal T}_{1}^{-}(\lambda_{1})\
\bar R_{2 1}(\lambda_{1}+\lambda_{2})\ {\cal T}_{2}^{-}(\lambda_{2}) \non \\
= {\cal T}_{2}^{-}(\lambda_{2})\ \bar R_{12}(\lambda_{1}+\lambda_{2})\
{\cal T}_{1}^{-}(\lambda_{1})\ R_{21}(\lambda_{1}-\lambda_{2}) \,.
\label{fundamental}
\ee
Now we are ready to prove (\ref{commutativity}),
\be
t(\lambda_{1})t(\lambda_{2}) &=& \tr_{1}  K_{1}^{+}(\lambda_{1})\ {\cal T}_{1}^{-}(\lambda_{1}) \tr_{2}  K_{2}^{+}(\lambda_{2})\ {\cal T}_{2}^{-}(\lambda_{2}) \non \\ &=& \tr_{1}  K_{1}^{+}(\lambda_{1})^{t_{1}}\ {\cal T}_{1}^{-}(\lambda_{1})^{t_{1}}\tr_{2}  K_{2}^{+}(\lambda_{2})\ {\cal T}_{2}^{-}(\lambda_{2}) \non \\ &=& \tr_{12}  K_{1}^{+}(\lambda_{1})^{t_{1}}\ K_{2}^{+}(\lambda_{2})\ {\cal T}_{1}^{-}(\lambda_{1})^{t_{1}}\  {\cal T}_{2}^{-}(\lambda_{2})\,,
\ee                                                                            we use the crossing unitarity of the $\bar R$ matrix namely,
\be                                                                            \bar R_{2 1}(-\lambda -2 \rho)^{t_{2}}\ \bar R_{2 1}(\lambda)^{t_{1}} = \zeta(\lambda)\,,
\ee
then the product $t(\lambda_{1})t(\lambda_{2})$ becomes,
\be
&&\zeta^{-1}(\lambda_{1}+\lambda_{2}) \tr_{12}   K_{1}^{+}(\lambda_{1})^{t_{1}}\ K_{2}^{+}(\lambda_{2})\ \bar R_{2 1}(-\lambda_{1}-\lambda_{2} -2 \rho)^{t_{2}} \bar R_{2 1}(\lambda_{1}+\lambda_{2})^{t_{1}}\ {\cal T}_{1}^{-}(\lambda_{1})^{t_{1}}\  {\cal T}_{2}^{-}(\lambda_{2}) \non \\&&
=  \zeta^{-1}(\lambda_{1}+\lambda_{2}) \tr_{12} (K_{1}^{+}(\lambda_{1})^{t_{1}}\ \bar R_{21}(-\lambda_{1}- \lambda_{2} -2 \rho)\ K_{2}^{+}(\lambda_{2})^{t_{2}})^{t_{2}} \non\\&& \times ({\cal T}_{1}^{-}(\lambda_{1})\ \bar R_{21}(\lambda_{1}+\lambda_{2})\ {\cal T}_{2}^{-}(\lambda_{2}))^{t_{1}} \non\\&& = \zeta^{-1}(\lambda_{1}+\lambda_{2}) \tr_{12} (K_{1}^{+}(\lambda_{1})^{t_{1}}\ \bar R_{2 1}(-\lambda_{1} -\lambda_{2} -2 \rho)\ K_{2}^{+}(\lambda_{2})^{t_{2}})^{t_{12}}
\non\\&& \times {\cal T}_{1}^{-}(\lambda_{1})\ \bar R_{2 1}(\lambda_{1}+\lambda_{2}) {\cal T}_{2}^{-}(\lambda_{2})\,,
\ee                                                                             using the unitarity of the $R$ matrix i.e.,
\be
R_{21}(-\lambda)\ R_{12}(\lambda) = \zeta(\lambda)\,,
\ee
we obtain the following expression for the product 
\be
&&\zeta^{-1}(\lambda_{1}-\lambda_{2})\zeta^{-1}(\lambda_{1}+\lambda_{2}) \tr_{12} (K_{1}^{+}(\lambda_{1})^{t_{1}}\ \bar R_{2 1}(-\lambda_{1} -\lambda_{2} -2 \rho)\ K_{2}^{+}(\lambda_{2})^{t_{2}})^{t_{12}}\non \\&& \times R_{21}(-\lambda_{1}+\lambda_{2})\ R_{12}(\lambda_{1}-\lambda_{2}) {\cal T}_{1}^{-}(\lambda_{1})\ \bar R_{2 1}(\lambda_{1}+\lambda_{2})\ {\cal T}_{2}^{-}(\lambda_{2}) \non \\ &&=  \zeta^{-1}(\lambda_{1}-\lambda_{2}) \zeta^{-1}(\lambda_{1}+\lambda_{2}) \tr_{12} (R_{12}(-\lambda_{1} + \lambda_{2})\ K_{1}^{+}(\lambda_{1})^{t_{1}}\  \bar R_{2 1}(-\lambda_{1}-\lambda_{2} -2 \rho)\ K_{2}^{+}(\lambda_{2})^{t_{2}})^{t_{12}}\non \\&& \times R_{12}(\lambda_{1}-\lambda_{2})\ {\cal T}_{1}^{-}(\lambda_{1})\ \bar R_{2 1}(\lambda_{1} + \lambda_{2} )\ {\cal T}_{2}^{-}(\lambda_{2})\,.
\ee                                                                            Finally, with the help of equations (\ref{boundaryYB2}) and (\ref{fundamental}) and repeating all the previous steps in a reverse order we end up that the the last expression is just $t(\lambda_{2})t(\lambda_{1})$. In order to show (\ref{commutativityp}) we need to define the following operator by changing the auxiliary space to its conjugate in (\ref{oper})
\be
{\cal T}_{\bar 0}^{-}(\lambda) = T_{\bar 0}(\lambda)K_{ \bar 0}^{-}(\lambda) \hat T_{0}(\lambda)\,.
\ee                                                              
${\cal T}_{\bar 0}^{-}(\lambda)$ satisfies the same fundamental relation (\ref{fundamental}) with ${\cal T}_{0}^{-}(\lambda)$ (remember (\ref{note})). Following exactly the same steps as before we can show that (\ref{commutativityp}) is also true.

It is also necessary to prove (\ref{commutativity3}). The steps of the proof are very similar to the previous case. The only difference is that this time we have to consider the reflection equation (\ref{boundaryYB1p}).                  Using (\ref{boundaryYB1p}) we can show the fundamental relation for  ${\cal T}_{\bar 0}^{-}(\lambda)$ and ${\cal T}_{0}^{-}(\lambda)$,
\be
\bar R_{1 2}(\lambda_{1}-\lambda_{2})\ {\cal T}_{\bar 1}^{-}(\lambda_{1})\
R_{21}(\lambda_{1}+\lambda_{2})\ {\cal T}_{2}^{-}(\lambda_{2}) \non \\
= {\cal T}_{2}^{-}(\lambda_{2})\ R_{12}(\lambda_{1}+\lambda_{2})\
{\cal T}_{\bar 1}^{-}(\lambda_{1})\ \bar R_{2 1}(\lambda_{1}-\lambda_{2}) \,.
\label{fundamental2}
\ee                                                                            Following the same procedure as before and using the relations       
\be                                                                            R_{21}(-\lambda -2 \rho)^{t_{2}}\ R_{21}(\lambda)^{t_{1}} = \zeta'(\lambda)\,,
\ee                                                                            and                                                                            \be
\bar R_{2 1}(-\lambda)\ \bar R_{1 2}(\lambda) = \zeta'(\lambda)
\ee                                                                            we can prove (\ref{commutativity3}). This concludes our proof for the integrability of the model.

\section{Appendix}
In this appendix we show explicitly that the Hamiltonian of the open spin chain is local with terms that describe interaction up to four nearest neighbours. We focus here in the special case that $K^{\pm}(\lambda) = 1$. We exploit the fact that $R_{ij}(0)={\cal P}_{ij}$, then the transfer matrices become (we write for simplicity  $\bar R_{ij}(0)= \bar R_{ij}$)     
\be
t(0) = \tr_{0}{\cal P}_{0 2N} \bar R_{0 2N-1}\cdots {\cal P}_{0 2k} \bar R_{0 2k-1}\cdots {\cal P}_{0 2} \bar R_{0 1 } \,, \non\\ \bar t(0) = \tr_{0} \bar R_{02N} {\cal P}_{0 2N-1}\cdots \bar R_{02k} {\cal P}_{0 2k-1}\cdots \bar R_{0 2} {\cal P}_{0 1 }\,.
\ee                                                                            We move the permutation operators along the elements of the product and having in mind that 
\be
\tr_{0}{\cal P}_{0 2N} \bar R_{0 2N}\ \propto 1\,, \qquad {\cal P}_{ij} A_{ik}{\cal P}_{ij} = A_{jk}\,,
\label{prop}
\ee
where $A$ is any operator, we end up
\be
t(0) \propto \bar R_{2N-1\ 2N} \bar R_{2N-3\ 2N-2} \cdots\bar R_{2k-1\ 2k} \cdots \bar R_{12} {\cal P}_{24}\cdots {\cal P}_{2k-2\ 2k}\cdots{\cal P}_{2N-2\ 2N} 
 \non\\ \times {\cal P}_{2N-3\ 2N-1}\cdots{\cal P}_{2k-3\ 2k-1}\cdots {\cal P}_{13} \bar R_{23}\cdots \bar R_{2k-2\ 2k-1}\cdots \bar R_{2N-2\ 2N-1}{\cal P}_{1\ 2N}\,,
\label{t}
\ee                      
and
\be
\bar t(0) \propto {\cal P}_{1\ 2N}\bar R_{2N-2\ 2N-1}\cdots \bar R_{2k-2\ 2k-1}\cdots\bar R_{23}{\cal P}_{13}\cdots{\cal P}_{2k-3\ 2k-1}\cdots {\cal P}_{2N-3\ 2N-1} \non\\ \times {\cal P}_{2N-2\ 2N}\cdots{\cal P}_{2k-2\ 2k}\cdots{\cal P}_{24}\bar R_{12}\cdots\bar R_{2j-1\ 2j} \cdots
\bar R_{2N-3\ 2N-2}\bar R_{2N-1\ 2N}\,.
\label{tbar}
\ee
We also need the derivative of the transfer matrix for $\lambda =0$. It is sufficient to show the calculation for the ${d\over d \lambda}t(\lambda)\bar t(\lambda)$, (the product  $t(\lambda){d\over d \lambda}\bar t(\lambda)$ gives similar terms). Taking the derivative of the transfer matrix we obtain four different sums, because the derivative hits $R$, $\bar R$ of the monodromy matrix $T$ and $\hat T$ as well, namely
\be
&&{d \over d \lambda}t(\lambda)\Big \vert_{\lambda=0} = \sum_{j=1}^{N} \tr_{0} {\cal P}_{0 2N} \bar R_{0 2N-1}\cdots \bar R'_{0 2j-1} \cdots {\cal P}_{0 2} \bar R_{01}  {\cal P}_{10} \bar R_{20} \cdots {\cal P}_{0 2N-1}  \bar R_{2N 0} \non\\&&
+ \sum_{j=1}^{N} \tr_{0} {\cal P}_{0 2N} \bar R_{0 2N-1} \cdots R'_{0 2j} \cdots \bar R_{01}  {\cal P}_{10} \bar R_{20}\cdots {\cal P}_{0 2N-1} \bar R_{2N 0}\non\\&&
+ \sum_{j=1}^{N} \tr_{0} {\cal P}_{0 2N} \bar R_{0 2N-1} \cdots {\cal P}_{0 2} \bar R_{01}  {\cal P}_{10} \bar R_{20}\cdots \bar R'_{0 2j} \cdots  {\cal P}_{0 2N-1} \bar R_{2N 0}\non\\&&
+ \sum_{j=1}^{N} \tr_{0} {\cal P}_{0 2N} \bar R_{0 2N-1}  \cdots {\cal P}_{0 2} \bar R_{01}  {\cal P}_{10} \bar R_{20}\cdots R'_{0 2j-1} \cdots {\cal P}_{0 2N-1}   \bar R_{2N 0}\,,
\ee                  
the prime denotes derivative with respect to $\lambda$. Again we move the permutation operators properly along the tensor product and we also consider (\ref{prop}) and $\tr_{0}{\cal P}_{0 2N} \bar R'_{0 2N} \propto 1$ and finally, we obtain                
\be
&&{d \over d \lambda}t(\lambda)\Big \vert_{\lambda=0} \propto  \sum_{j=1}^{N} \bar R_{2N-1\ 2N} \bar R_{2N-3\ 2N-2} \cdots\bar R'_{2j-1\ 2j} \cdots \bar R_{12} {\cal P}_{24}
\cdots {\cal P}_{2N-2\ 2N} \non\\&& \times {\cal P}_{2N-3\ 2N-1}\cdots {\cal P}_{13} \bar R_{23}\cdots \bar R_{2k-2\ 2k-1}\cdots \bar R_{2N-2\ 2N-1}{\cal P}_{1\ 2N} \non\\&&                  
+ \sum_{j=1}^{N-1} \bar R_{2N-1\ 2N} \bar R_{2N-3\ 2N-2} \cdots\bar R_{2j+1\ 2j+2} \check R'_{2j\ 2j+2} \bar R_{2j-1\ 2j} \cdots \bar R_{12} {\cal P}_{24}
\cdots {\cal P}_{2N-2\ 2N} \non\\&& \times {\cal P}_{2N-3\ 2N-1}\cdots {\cal P}_{13} \bar R_{23}\cdots \bar R_{2k-2\ 2k-1}\cdots \bar R_{2N-2\ 2N-1}{\cal P}_{1\ 2N} \non\\&&              
+\sum_{j=1}^{N-1} \bar R_{2N-1\ 2N} \bar R_{2N-3\ 2N-2} \cdots\bar R_{2k-1\ 2k} \cdots \bar R_{12} {\cal P}_{24}
\cdots {\cal P}_{2N-2\ 2N} \non\\&& \times {\cal P}_{2N-3\ 2N-1}\cdots {\cal P}_{13} \bar R_{23}\cdots \bar R'_{2j\ 2j+1}\cdots \bar R_{2N-2\ 2N-1}{\cal P}_{1\ 2N} \non\\ &&              
+ \sum_{j=1}^{N-2} \bar R_{2N-1\ 2N} \bar R_{2N-3\ 2N-2} \cdots \bar R_{2k-1\ 2k} \cdots \bar R_{12} {\cal P}_{24}
\cdots {\cal P}_{2N-2\ 2N} \non\\&& \times {\cal P}_{2N-3\ 2N-1}\cdots {\cal P}_{13} \bar R_{23}\cdots \bar R_{2j\ 2j+1} \check R'_{2j+1\ 2j+3} \bar R_{2j+2\ 2j+3}\cdots \bar R_{2N-2\ 2N-1}{\cal P}_{1\ 2N}
 \non\\&& + \tr_{0} \check R'_{0 2N} \bar R_{2N-1\ 2N} \bar R_{2N-3\ 2N-2} \cdots \bar R_{2k-1\ 2k} \cdots \bar R_{12} {\cal P}_{24}
\cdots {\cal P}_{2N-2\ 2N} \non\\&& \times {\cal P}_{2N-3\ 2N-1} \cdots {\cal P}_{13} \bar R_{23}\cdots \bar R_{2k-2\ 2k-1} \cdots \bar R_{2N-2\ 2N-1}{\cal P}_{1\ 2N} {\cal P}_{02N} {\bar R_{0 2N}}  + t(0) \non\\&& + \bar R_{2N-1\ 2N} \bar R_{2N-3\ 2N-2} \cdots\bar R_{2k-1\ 2k} \cdots \bar R_{12} {\cal P}_{24}
\cdots{\cal P}_{2N-2\ 2N} \non\\&& \times {\cal P}_{2N-3\ 2N-1} \cdots {\cal P}_{13} \check R'_{3\ 2N} \bar R_{23}\cdots \bar R_{2k-2\ 2k-1}\cdots \bar R_{2N-2\ 2N-1}{\cal P}_{1\ 2N} \non\\ && + \bar R_{2N-1\ 2N} \bar R_{2N-3\ 2N-2} \cdots\bar R_{2k-1\ 2k} \cdots \bar R_{12} {\cal P}_{24}\ \cdots{\cal P}_{2N-2\ 2N} \non\\ && \times 
 {\cal P}_{2N-3\ 2N-1} \cdots {\cal P}_{13} \bar R_{23}\cdots \bar R_{2k-2\ 2k-1}\cdots \bar R_{2N-2\ 2N-1} \check R'_{1\ 2N-1}{\cal P}_{1\ 2N}\,,            
\label{der}
\ee
where $\check R'_{ij} = {\cal P}_{ij} R'_{ij}$. The last four terms in (\ref{der}) come from the second, and the third sum for $j=N$ and from the last sum for $j=0$ and $j=N-1$, respectively. Combining (\ref{tbar}), (\ref{der}) and having in mind that ${\cal P}_{ij} A_{ik}{\cal P}_{ij} = A_{jk}$ and ${\cal P}_{ij}^{2}, \bar R_{ij}^{2} \propto 1$,  we get
\be
&&{d \over d \lambda}t(\lambda) \bar t(\lambda) \Big \vert_{\lambda=0} \propto  \sum_{j=1}^{N}
\bar R'_{2j-1\ 2j} \bar R_{2j-1\ 2j}              
+ \sum_{j=1}^{N-1} \bar R_{2j+1\ 2j+2} \check R'_{2j\ 2j+2} \bar R_{2j+1\ 2j+2} \non\\&&               
+ \sum_{j=1}^{N-1} \bar R_{2j+1\ 2j+2} \bar R_{2j-1\ 2j} \bar R'_{2j-1\ 2j+2}\bar R_{2j-1\ 2j+2}\bar R_{2j-1\ 2j}\bar R_{2j+1\ 2j+2} \non\\&&               
+ \sum_{j=1}^{N-1} \bar R_{2j+1\ 2j+2} \bar R_{2j-1\ 2j} \bar R_{2j-1\ 2j+2} \check R'_{2j-1\ 2j+1} \bar R_{2j-1\ 2j+2} \bar R_{2j-1\ 2j}\bar R_{2j+1\ 2j+2}\non\\&&
+ \tr_{0} \check R'_{0 2N} \bar R_{2N-1\ 2N} {\cal P}_{0 2N-1} \bar R_{0 2N-1} \bar R_{2N-1\ 2N} + t(0) \bar t(0) + \bar R_{12} \check R'_{12}\bar R_{12}\,.
\label{terms}
\ee
Notice that the first two terms of the last equation give exactly the Hamiltonian of the alternating spin chain constructed by De Vega and Woyanorovich (see e.g. \cite{VEWO}, \cite{abad}). The last term of (\ref{der}) is included in the fourth sum of (\ref{terms}) for $j=N-1$.  It is obvious from (\ref{t}), (\ref{tbar}) that $t(0) \bar t(0) \propto 1$. Equation (\ref{terms}) contains all the Hamiltonian's terms (remember $ t(\lambda){d \over d\lambda}\bar t(\lambda)$ has a similar form to (\ref{terms})). We observe that the terms in (\ref{terms}) describe local interaction between two, three and four nearest neighbours. We conclude that this is indeed a local Hamiltonian.

\end{document}